\documentclass{aa}  

\usepackage{graphicx}

\usepackage{txfonts}
\usepackage{threeparttable}
\usepackage{orcidlink}

\newcommand{\FIG}[1]{Fig.~\ref{fig:#1}}
\newcommand{\TAB}[1]{Table~\ref{tab:#1}}
\newcommand{\SEC}[1]{Section~\ref{sec:#1}}

\begin{document}

   \title{An RFSoC-based F-engine for ARGOS}

\author{Yunpeng Men\orcidlink{0000-0003-4137-4247}\inst{1}
        \and Ewan Barr\orcidlink{0000-0001-8715-9628}\inst{1}
        \and Amit Bansod\inst{1}
        \and Weiwei Chen\orcidlink{0000-0002-6089-7943}\inst{1}
        \and Jason Wu\inst{1}
        \and John Antoniadis\orcidlink{0000-0003-4453-3776}\inst{1,2}
        \and \\ Jan Behrend\orcidlink{0000-0002-8622-1298}\inst{1}
        \and Niclas Esser\orcidlink{0009-0001-3328-1895}\inst{1}
        \and Oliver Polch\inst{1}
        \and Gundolf Wieching\orcidlink{0000-0002-2758-1840}\inst{1}
        \and Tobias Winchen\orcidlink{0000-0002-5756-4304}\inst{1}
        }

\institute{
    Max-Planck-Institut f\"ur Radioastronomie, Auf dem H\"ugel 69, D-53121 Bonn, Germany\\
    \email{ypmen@mpifr-bonn.mpg.de}
    \and
    Institute of Astrophysics, Foundation for Research and Technology Hellas, N. Plastira 100, 70013, Heraklion, Greece
}

   \date{Received XX XX, XXXX; accepted XX XX, XXXX}

 
  \abstract
 {Radio interferometers provide the means to perform the wide field-of-view (FoV) high-sensitivity observations required for modern radio surveys. As computing power per cost has increased, there has been a move toward larger arrays of smaller dishes, such as DSA-2000, the upcoming HIRAX, CHORD, and SKA radio telescopes. Such arrays can have simpler receiver designs with room-temperature low-noise amplifiers and use direct sampling to greatly reduce the cost per antenna. The ARGOS project is currently developing an array of five six-meter antennas that will be used to demonstrate the technology required for a next-generation ``small-D, big-N'' radio interferometer in Europe.}
   {For this work our objective was to implement a first-stage digital signal processing system for the ARGOS demonstrator array, providing digitization, channelization, delay correction, and frequency-dependent complex gain correction. The system is intended to produce delay- and phase-corrected dual-polarization channelized voltages in the frequency range 1-3 GHz with a nominal channel bandwidth of 1 MHz.}
   {We used a Radio Frequency System-on-Chip (RFSoC) 4x2 evaluation board with four analog-to-digital converters (ADCs) that can simultaneously sample two 1\,GHz dual-polarization bands. A critically sampled polyphase filter bank (PFB) using an 8-tap finite impulse response (FIR) filter and a 2048-point fast Fourier transform (FFT) was applied to channelize the input data. Coarse and fine delays were corrected separately before and after the PFB. The post-PFB data were gain corrected before a corner-turner was applied to transpose the channelized data into time-minor order for efficient network transmission. The data were packetized and transmitted over a 100-GbE network. We used Xilinx Vitis HLS C++ to develop the required firmware as a set of customizable modules suitable for rapid prototyping.}
   {We performed hardware verification of the channel response of the critically sampled PFB and of the delay correction, showing  both to be consistent with theoretical expectations. Furthermore, the board was installed at the Effelsberg 100-meter radio telescope where we performed commensal pulsar observations with the Effelsberg Direct Digitization backend, showing comparable performance. This work demonstrates the utility of high-level synthesis (HLS) languages in the development of high-performance radio astronomy processing backends.}   
   {}

\titlerunning{A RFSoC-based F-engine for ARGOS}
 \authorrunning{Men et al.}
   \maketitle
%

\section{Introduction}
The radio sky exhibits dynamic characteristics;  various objects emit transient radio signals, such as fast radio bursts \citep{CHIME2021ApJS, Rajwade2022MNRAS, Shannon2025PASA}, radio pulsars \citep{Cordes2006ApJ, Keith2010MNRAS,Li2018IMMag, Han2021RAA, Padmanabh2023MNRAS}, and slow radio transients \citep{Hurley-Walker2022Nat, Hurley-Walker2023Nat, Hurley-Walker2024ApJ, Caleb2024NatAs, Ruiter2024arXiv, Dong2024arXiv}. To improve the efficiency of capturing these radio signals, large fields of view and high sensitivities are required, motivating the concept of ``small-D, big-N'' radio interferometers such as MeerKAT \citep{Jonas2016mks} with tens of dishes, and upcoming arrays with thousands of dishes, including HIRAX \citep{Crichton2022JATIS}, CHORD \citep{Vanderlinde2019clrp}, and DSA-2000 \citep{Hallinan2019BAAS}. The development of these interferometers is enabled by the rapid decrease in the price of computing power, making affordable the large processing backends required to analyze the data they produce.

The ARGOS project is a concept for a next-generation, low-cost, sustainable, small-D, big-N radio interferometer to be located in Europe. To demonstrate this concept, an array of five 6-meter radio telescopes are under construction on the Greek island of Crete. The telescopes will use uncooled receivers covering the frequency band from 1 to 3\,GHz, the signals from which will be directly sampled, thus eliminating the need for heterodyning. Thanks to its simplicity, this  front-end design is intended to significantly reduce the cost per antenna. The back-end system will include a channelizer capable of producing 1 MHz resolution dual-polarization channelized voltages; a correlator for producing time-averaged visibilities for all array baselines; a beamformer to generate full-Stokes synthesized beams for all antennas and frequency channels; a pulsar timing subsystem for performing pulsar folding and automatic timing analyses; and a fast transient detection subsystem. ARGOS will further include a dedicated imaging subsystem as well as robust real-time anomaly detection and alerting systems. The details of the ARGOS project will be presented in a future paper (Antoniadis et al., in prep.).

In this work we present the development of the first stage digital processing system for the ARGOS demonstrator, implementing digitization, channelization, and delay and gain correction. To process the 1-3\,GHz frequency band required by ARGOS, we made use of the Radio Frequency System-on-Chip  (RFSoC) 4x2 board produced by Real Digital.\footnote{\url{https://www.realdigital.org/hardware/rfsoc-4x2}} The board uses AMD’s ZYNQ Ultrascale+ RFSoC technology, which integrates high-speed analog-to-digital converters (ADCs), digital-to-analog  converters (DACs), a field-programmable gate array (FPGA), and embedded ARM processors onto a single chip. The RFSoC 4x2 board uses the Gen 3 ZU48DR chip with four radio frequency (RF) inputs, each of which can be digitized at sampling rates up to 5 GSPS. To directly digitize the required 1-3\,GHz band, the input signal will be divided into 1-2\,GHz and 2-3\,GHz bands by two bandpass filters before being fed to the RF inputs. The four inputs of the board are then configured to sample at 2\,GSPS resulting in each polarization being sampled in both the second and third Nyquist zones. Channelized data are quantized to (8,8) bit complex samples before being transmitted out of the board via a QSFP28 interface supporting 100 GbE. 

Field-programmable gate array development is a multi-stage process involving design, simulation, and hardware testing. Traditionally, hardware description languages (HDLs) such as Verilog and VHDL are used to describe digital circuits at the register-transfer level (RTL). While powerful for precise hardware control, HDLs are less suited for algorithmic descriptions, making rapid prototyping challenging. Moreover, development with HDLs often emphasizes RTL simulations, which can be time-consuming and less intuitive for algorithm verification. To overcome these limitations, high-level synthesis (HLS) languages have gained popularity \citep{CongIEEE2011}. HLS allows designers to describe systems at an algorithmic level, enabling faster prototyping and abstracting away low-level timing details during early design phases. It also supports high-level simulations prior to RTL simulation, accelerating the verification process. For this work we used Xilinx Vitis HLS C++ to implement multiple components of the ARGOS digitizer/channelizer. 

In this paper we present the top-level design of the ARGOS digitizer/channelizer and the implementation of its components in \SEC{Methods}. The benchmarks and experiments with real observations are presented in \SEC{Benchmark}. We discuss the experience of using HLS and future improvements in \SEC{Discussion}, and summarize our conclusions in \SEC{Conclusions}.

\section{Methods}
\label{sec:Methods}

\begin{figure*}
    \centering
    \includegraphics[width=1.7\columnwidth]{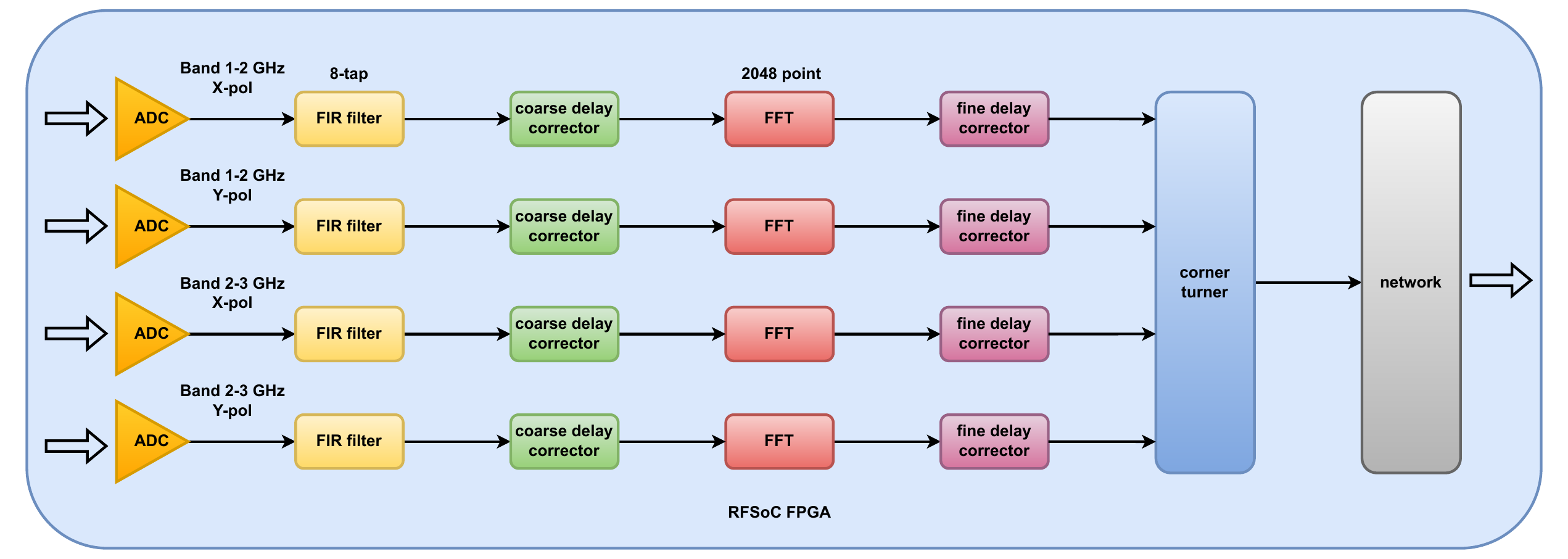}
    \caption{Dataflow diagram of the F-engine design. The dual-band and dual-polarization voltages are digitized by ADCs and then processed sequentially by the 8-tap FIR filter, coarse delay corrector, FFT block, and fine delay corrector. The four data streams are then merged and transposed in the corner turner before being packetized and transferred through the 100G network interface.}
    \label{fig:top_design}
\end{figure*}

In an FX correlator, the F-stage (or F-engine) performs three primary functions: channelization, delay correction, and complex gain correction. Channelization enables frequency-dependent signal correction, efficient data multiplexing across the correlator hardware, and the cross-correlation of antenna pairs through complex multiplication.

Delay correction ensures the  time alignment of received signals from all antennas. It accounts for  geometric delays--determined by the array’s configuration and pointing direction--and for instrumental delays arising from path length differences between antennas, the correlator, and the observatory clock. The total delay is decomposed into an integer component (coarse delay), corrected before channelization, and a fractional component (fine delay), corrected by applying a frequency-dependent phase ramp to the channelized output. During this fine delay correction, additional phase and amplitude adjustments may be applied to compensate for gain variations across antennas, polarizations, and frequencies.

To realize these functions we developed a modular streaming pipeline design, shown in \FIG{top_design}. The data flow can be described as follows: (1) The four RF inputs are digitized by the ADC modules; (2) the digitized data are then shifted by integer sample delays through the coarse delay correctors; (3) the data are channelized by the polyphase filter bank modules (PFBs) which use an 8-tap polyphase filter with a windowing function and a fast Fourier transform (FFT); (4) the channelized data are multiplied by complex vectors, correcting for both the residual fine delay of the input data and complex gain; (5) after delay correction and channelization, the data are buffered for multiple samples and transposed (corner-turned) between the time and frequency dimensions; (6) the transposed data are then transferred through the network module in user datagram protocol (UDP) packets, with one frequency channel and one polarization per packet. The detailed design of each module is described in the following sections.

\subsection{Polyphase filter bank}
\begin{figure}
    \centering
    \includegraphics[width=\columnwidth]{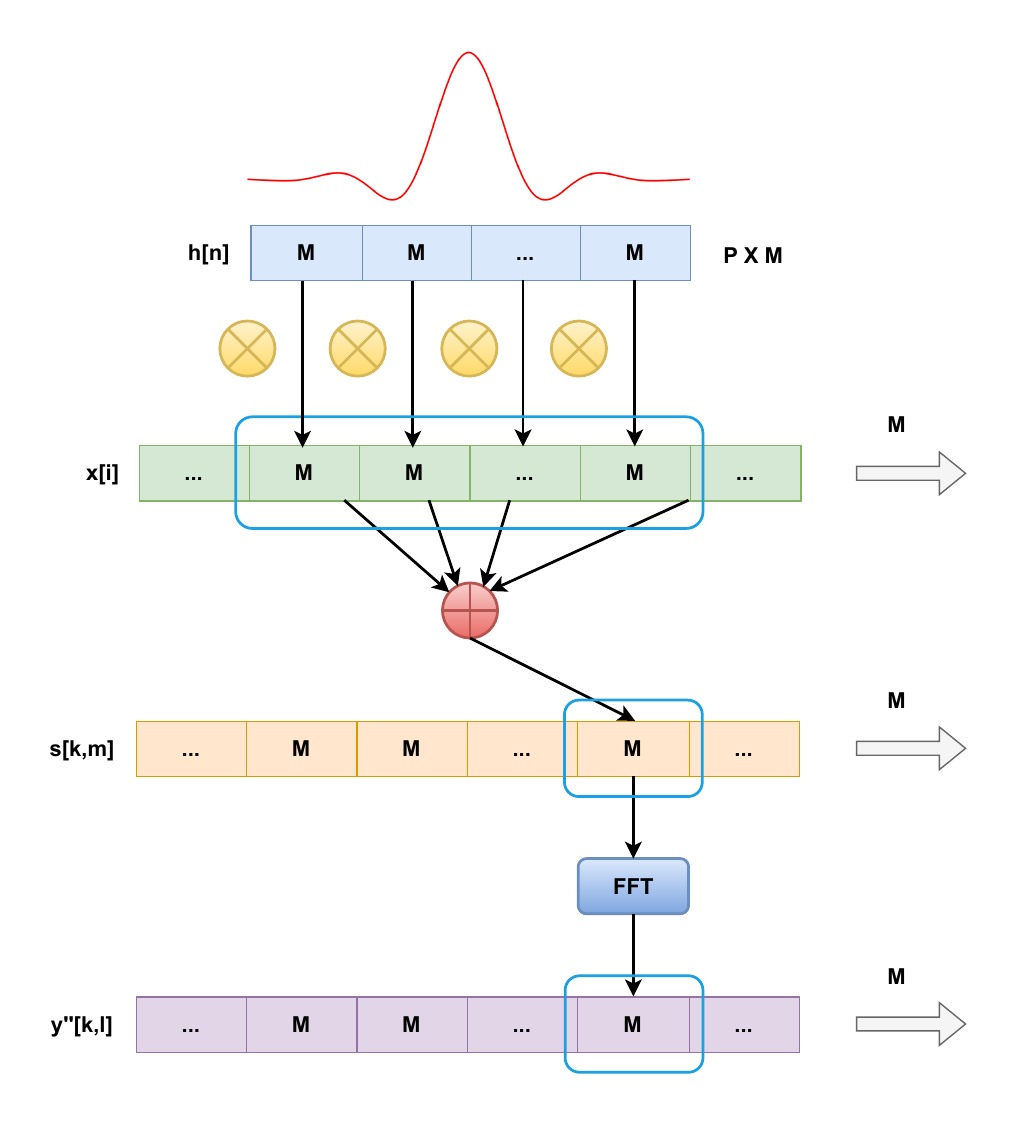}
    \caption{Diagram illustrating the algorithm of a critically sampled polyphase filterbank (PFB). The input stream is segmented into frames of $M$ samples. At each step, a window of $P$ consecutive frames is multiplied by predefined filter coefficients, advanced by one frame per iteration. The filtered outputs are summed across the $P$ frames to yield a single $M$-sample frame, which is then transformed via FFT.}
    \label{fig:pfb}
\end{figure}

Spectral leakage, wherein signal power spreads into neighboring frequency channels instead of remaining confined to its true frequency, leads to unwanted spectral contamination, particularly at channel boundaries. In astronomical observations, minimizing this leakage is critical as it can degrade calibration accuracy, introduce spectral contamination from strong narrowband interference, reduce beamforming coherence by distorting frequency-dependent delay corrections, and broaden dispersed time-domain signals, impacting pulsed and transient signal detection. To suppress the spectral leakage, we channelized using PFBs \citep{Vaidyanathan1993}. These provide good channel-to-channel isolation by incorporating a multi-tap finite impulse response (FIR) filter prior to application of a Fourier transform. The output $\pmb{y}$ of a signal $\pmb{x}$ passing through a FIR filter can be described as the convolution between the input signal and a vector of filter coefficients $\pmb{h}$, i.e.,
\begin{equation}
    y[n] = \sum^{N-1}_{i=0} x[n-i] \cdot h[i] \,,
\end{equation}
where is $N$ is the number of filter coefficients. The corresponding frequency response is 
\begin{equation}
    Y(\omega) = X(\omega) \cdot H(\omega)\,,
\end{equation}
where $Y(\omega)$, $X(\omega)$, and $H(\omega)$ are the Fourier amplitudes at a frequency $\omega$ of $x[i]$, $y[n]$, and $h[i]$, respectively, which have normalized units of radians/sample. To obtain multiple frequency channels, a bank of FIR filters is applied; the output of the $l$-th filter is
\begin{equation}
    y'[n, l] = \sum^{N-1}_{i=0} x[n-i] \cdot h[i]\ \exp \left(j 2\pi \frac{l}{M} i \right) \,,
\end{equation}
where $M$ is the number of frequency channels. The corresponding frequency response at $l$-th channel can be given as
\begin{equation}
    Y(\omega, l) = X(\omega) \cdot H(\omega-2\pi\frac{l}{M})\,.
\end{equation}
Because the channel width is $2\pi/M$, we can perform decimation by a factor of $M$ to reduce the output data rate. Applying a $P$-tap FIR filter with $P\times M$ coefficients, the output of the $l$-th channel can be given as
\begin{align}
    y''[k, l] &= y'[k M, l] \notag\\
    &= \sum^{M-1}_{m=0} s[k, m] \exp \left(j 2\pi \frac{l}{M} m \right)\,,
    \label{eq:pfb}
\end{align}
with
\begin{equation}
    s[k, m] = \sum^{P-1}_{p=0} x[k M-(p M + m)] \cdot h[p M + m]\,,
\end{equation}
where $y''[k, l]$ represents the $k$-th sample at $l$-th channel. The output data are produced by applying a $P$-tap FIR filter and FFT on the input data, as shown in \FIG{pfb}. In our work, we used an 8-tap FIR filter with a Hamming window, the frequency response is shown in \FIG{channel_response}. The Hamming window is adopted to effectively suppress sidelobes, while maintaining simple implementation. An 8-tap design is used to balance between achieving an optimal spectral response and minimizing FPGA resource usage. Since we aim for better leakage suppression, the current PFB design cannot be perfectly inverted; however, this limitation has little impact on most radio astronomy applications. In future work, we may consider implementing an oversampled PFB (see \SEC{Discussion}), which provides improved invertibility.

The FIR filter was implemented in the Xilinx Vitis HLS C++ language. The input and output have data widths of 16 bits, while the coefficients of the FIR filter were stored in the Block RAM on FPGA, which has a width of 27 bits with a 2 bit integer that can fit the DSP48 resources on FPGA. The filter can process eight samples per clock  cycle to match the input rate from the ADCs. The Xilinx Vector FFT in the Xilinx Super Sample Rate (SSR) blockset of Vitis Model Composer is used to perform the FFT. This implementation is capable of processing multiple samples per clock cycle at a clock rate of 500\,MHz, which meets our performance requirements. In our design, we use the FFT size of 2048, SSR of 8 for each ADC input. The FFT output has 27 bits, including 14 integer bits without scaling. Therefore, for a 14-bit input and a 2048-point FFT, there is no loss of precision or dynamic range in the FFT computation.

\subsection{Geometrical delay correction and complex gain calibration}
In array signal processing, it is necessary to correct for the geometrical delay arising from the spatial offset between each antenna element and the array reference point. Failure to apply accurate delay correction degrades the array’s directional gain. Delay correction consists of both time-domain and frequency-domain components; integer-sample delays are corrected in the time domain, while subsample delays and frequency-dependent complex gain are compensated for in the frequency domain. The geometrical delay is computed using a first-order polynomial model, with coefficients periodically updated via communication between the CPU and the firmware. In our implementation, integer and subsample delay corrections are performed by the coarse and fine delay correctors, respectively. Frequency-dependent complex gain calibration is also handled by the fine delay corrector. Both delay correctors are implemented using Xilinx Vitis HLS in C++.

\begin{figure}
    \centering
    \includegraphics[width=\columnwidth]{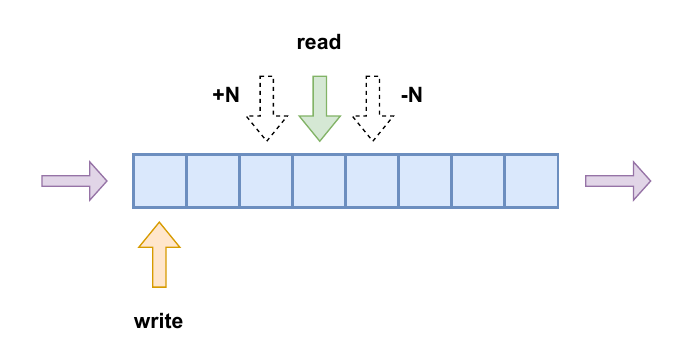}
    \caption{Diagram illustrating the algorithm of the coarse delay correction. The input data stream is continually written in a ring buffer, while the buffer is read into the output data stream with a shift starting position based on integer delay setup, which can be updated on the fly.}
    \label{fig:coarse_delay}
\end{figure}

\paragraph{Coarse delay corrector} To enable real-time adjustment of integer delays, we implemented an indexed buffer, where the data are written sequentially and read based on the configured integer delay, as illustrated in \FIG{coarse_delay}. The buffer is implemented using block RAM (BRAM) on the FPGA, supporting a maximum integer delay of 65536 samples. At a sampling rate of 2\,GHz, this allows  a delay compensation of approximately 300\,$\mu$s, which corresponds to a maximum baseline of 90 km in a radio array. The integer delay is dynamically computed using a linear interpolation delay predictor, ensuring precise and efficient real-time correction.

\paragraph{Fine delay corrector} A fine delay corrector was implemented to compensate for subsample delay and complex gain variations. The corrector processes the PFB output, applying a complex phase correction to each frequency channel. The complex phase factors are derived from two components: (1) fractional geometrical delay terms are computed for each frequency channel using a linear interpolation delay predictor, which is updated in real time; (2) complex gain terms are stored in dual-port block RAM (BRAM) on the FPGA and can be set up through communication between the Programmable Logic (PL) and Processing System (PS) on the RFSoC.

\paragraph{Output rescaling} Due to the limitations of network bandwidth, we rescaled the output to the range of signed 8-bit integers after the complex gain multiplication applied in the fine delay corrector. The complex gain consists of a 16-bit amplitude and a 16-bit phase component. Applying this gain reduces the dynamic range of the signal to 8 bits, allowing efficient data transmission while maintaining the essential amplitude and phase information.

\begin{figure*}
    \centering
    \includegraphics[width=1.8\columnwidth]{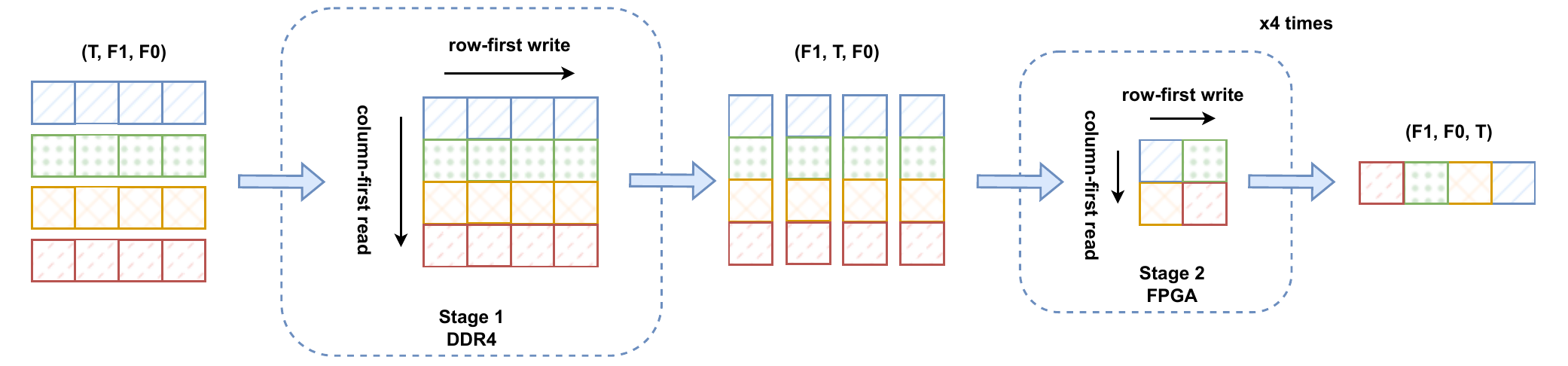}
    \caption{Diagram illustrating the algorithm of the corner turner. The corner turner performs transposition between time and frequency dimension in blocks of 512 spectra in two stages, with each stage transposing the data of fractional spectra.}
    \label{fig:corner_turner}
\end{figure*}

\subsection{Corner turner}
Beamforming and correlation require data from all antennas for each frequency channel. Given the high aggregate data rate from all antennas, the processing must be distributed across multiple computing nodes. This necessitates routing different frequency channels to different destinations, a function performed by the F-engine. In our design, the channelized output data are first accumulated over 512 samples, then transposed between the time and frequency dimensions. This transposition is achieved via indexed reads and writes to a buffer with a carefully designed access pattern. To optimize the throughput, a two-stage transposition scheme is employed using both the DDR4 memory on the RFSoC4x2 board and the FPGA’s UltraRAM. In the first stage, time-sequential spectral data are written to DDR4 and are then read out in a transposed access pattern, as illustrated in \FIG{corner_turner}. Because DDR4 throughput can degrade under nonsequential access, only a subset of frequency channels are transposed in this stage. The remaining channels are transposed in a second stage using a smaller buffer implemented in UltraRAM on the FPGA. As a result, the corner turner outputs data ordered as multiple time samples of a single frequency channel in contiguous blocks. These are subsequently packetized and transmitted over the network. The full corner turner system was implemented in Xilinx Vitis HLS using C++.

\subsection{Packetization}
The channelized data are packetized before being transmitted over the network to ensure that the receiving end can correctly interpret and reassemble the data stream. To facilitate this, we adopted the {\sc SPEAD2} protocol,\footnote{\url{https://github.com/ska-sa/spead2}} a high-performance streaming data format widely used in radio astronomy for real-time data transport. In our implementation, each data heap consists of a single packet to reduce latency and simplify parsing on the receiving side. The header of each heap includes two custom metadata fields: (1) the timestamp, representing the number of clock cycles elapsed since the start of data acquisition, which enables precise temporal alignment of data streams; (2) the ordering vector, which encodes the index information corresponding to the antenna ID, frequency channel, and polarization channel, which is essential for correctly reconstructing the multidimensional data array at the receiving side. The payload of each SPEAD packet contains only the raw channelized data, which together with the lightweight header maximizes bandwidth efficiency. This streamlined packet structure ensures low-latency high-throughput data transmission, while retaining sufficient metadata for downstream processing. Currently, each packet contains 512 complex 8-bit integers, resulting in a 1\,kB payload. While a larger payload size may be desirable for network efficiency reasons, the design is currently restricted by resource usage in the corner turner.

\subsection{Control and monitor}
In our F-engine design, we used the PS-PL interfaces on the RFSoC FPGA to perform control and monitoring of the data processing pipeline. The pipeline is started and stopped by setting up the corresponding registers. The starting of the pipeline is synchronized with the one-pulse-per-second (PPS) signal. When the stop signal is sent, the pipeline will stop and all registers and buffers will be reset. The pipeline can be restarted again without reloading the firmware. To monitor the data stream on the fly, we implemented a snapshot kernel using Xilinx Vitis HLS C++, which can capture one frame of output data of the ADC, coarse delay corrector, and fine delay corrector when a trigger signal is received from PS. This design provides the capability of on-the-fly debugging.

Although the RFSoC4x2 board includes a 1G network interface for control, we implemented a virtual 1G network interface over the existing 100G network interface to minimize the number of physical connections between the board and other devices. This was achieved in the FPGA firmware using Xilinx network IP. The 100G network provides sufficient bandwidth for both control and high-speed data transfer, ensuring efficient communication without additional hardware. Moreover, this approach simplifies the design of future custom FPGA boards by reducing the need for multiple network interfaces, thereby reducing design complexity.

\section{Tests and results}
\label{sec:Benchmark}
We characterized the performance of the F-engine, including resource usage, power consumption, and channel response. Additionally, a test observation was conducted on PSR\,J1939+2134. The details are presented below.

\subsection{Resource usage and power consumption}
To support programmable digital circuits for diverse use cases, an FPGA has various resources, including (1) look-up tables (LUTs), which can be used for combinational logic or distributed memory; (2) flip-flops (FFs), which are registers that can be used for timing logic; (3) block RAM (BRAM), which are memory units with configurable data width and depth; (4) ultra RAM (URAM); (5) digital signal processing slices (DSPs), which are specified hardware cores used for high-performance computing; and (6) other resources, such as clocks, I/Os, and transceivers. Optimizing resource usage can reduce power consumption and improve throughput. The Xilinx Vivado suite can provide the measurements of the resource usage and power consumption for a design, which can help to discover the bottlenecks and optimize the design to reduce the resource usage. \TAB{resource_usage} shows the statistics of the resource usage and power consumption on FPGA. The utilization of LUTs, FFs, BRAM, URAM, and DSP are 68\%, 42\%, 74\%, 40\%, and 26\%, respectively. As expected, the PFBs consume the majority of the logic resources. The total power consumption of the FPGA firmware is approximately 30 W. Additionally, it is important to note that other components on the RFSoC4x2 board contribute an additional power consumption of about 30 W.

   \begin{table*}
      \caption[]{Resource usage of the F-engine.}
      \centering
         \label{tab:resource_usage}
         \begin{tabular}{lrrrrrrr}
            \hline
            \noalign{\smallskip}
            Module      &  LUTs & FFs & BRAM & URAM & DSPs & Power (W) & DDR4 (GB/s)\\
            \noalign{\smallskip}
            \hline
            \noalign{\smallskip}
            FIR & 9676 & 4922 & 52 & 0 & 65 & 0.5 & --\\
            FFT & 18147 & 25554 & 4 & 0 & 63 & 2.4 & --\\
            coarse delay corrector & 556 & 911 & 32 & 0 & 0 & 0.1 & --\\
            fine delay corrector & 2570 & 8427 & 4 & 0 & 56 & 0.1 & --\\
            corner turner & 15566 & 21860 & 24 & 32 & 0 & 1.1 & 8.192\\
            network & 25089 & 33375 & 83 & 0 & 0 & 3.9 & --\\
            total\tablefootmark{a} & 260291 & 367971 & 765 & 32  & 750 & 30.0 & 8.192\\
            available &  425280 & 850560 & 1080 & 80 & 4272 & -- & 10.664\tablefootmark{b}\\
            percentage &  68\% &  42\% & 74\% & 40\% &  26\% & -- & 77\%\\
            \noalign{\smallskip}
            \hline
         \end{tabular}
    \tablefoot{
        The resource usage for the FIR, FFT, coarse delay corrector, and fine delay corrector is measured per data stream. Since there are four data streams in total, the overall resource consumption scales accordingly.\\
        \tablefoottext{a} The total resource usage and power consumption include all modules not listed in the table.
        \tablefoottext{b} The theoretical DDR4 bandwidth is based on a clock frequency of 1333\,MHz. However, experimental tests show a maximum achievable throughput of only 8.7\,GB/s.
   }
   \end{table*}

\subsection{Channel response}
\begin{figure}
    \centering
    \includegraphics[width=0.9\columnwidth]{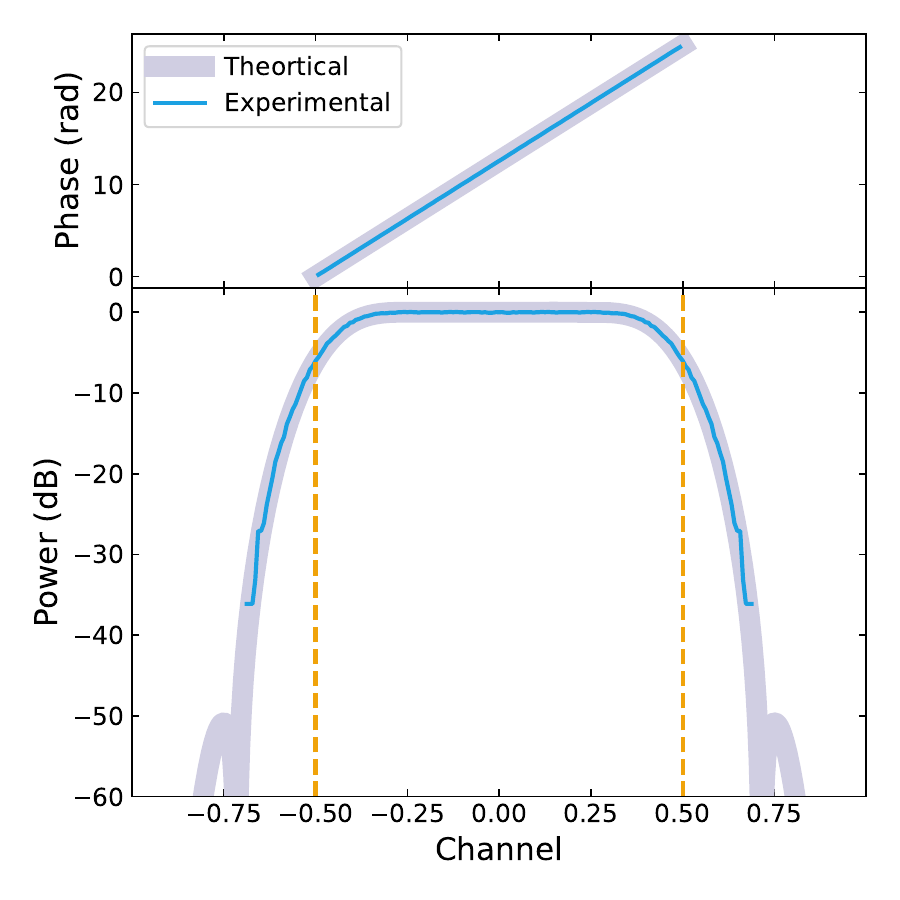}
    \caption{Measured channel frequency response of the critically sampled PFB. The upper panel shows the phase response and the bottom panel shows the amplitude response. The thin blue solid line represents the measured channel response, while the thick purple solid line represents the  theoretical channel response. The yellow dashed lines represent the edges of one channel.}
    \label{fig:channel_response}
\end{figure}

To verify the hardware functionality of the channelization process, we conducted measurements of the channel response of the PFB. Utilizing the integrated ARM CPUs on the Xilinx RFSoC chip, which supports a processing system, we simulated a sine wave to stimulate the input of the PFB kernel via the communication channel between PL and PS instead of using real data from the ADC inputs. This setup allowed us to input an ideal sweeping sine wave with a constant amplitude, enabling precise measurement of the channel response. We applied a frequency sweep ranging from 511\,MHz to 513\,MHz, which spans twice the channel bandwidth of the PFB. This sweep allowed us to thoroughly evaluate the channel response, focusing specifically on the frequency channel at 512\,MHz, as shown in \FIG{channel_response}. The results indicate that the measured channel response aligns well with the theoretical calculations, confirming the correct operation of the PFB.

\subsection{Delay correction}
\begin{figure}
    \centering
    \includegraphics[width=\columnwidth]{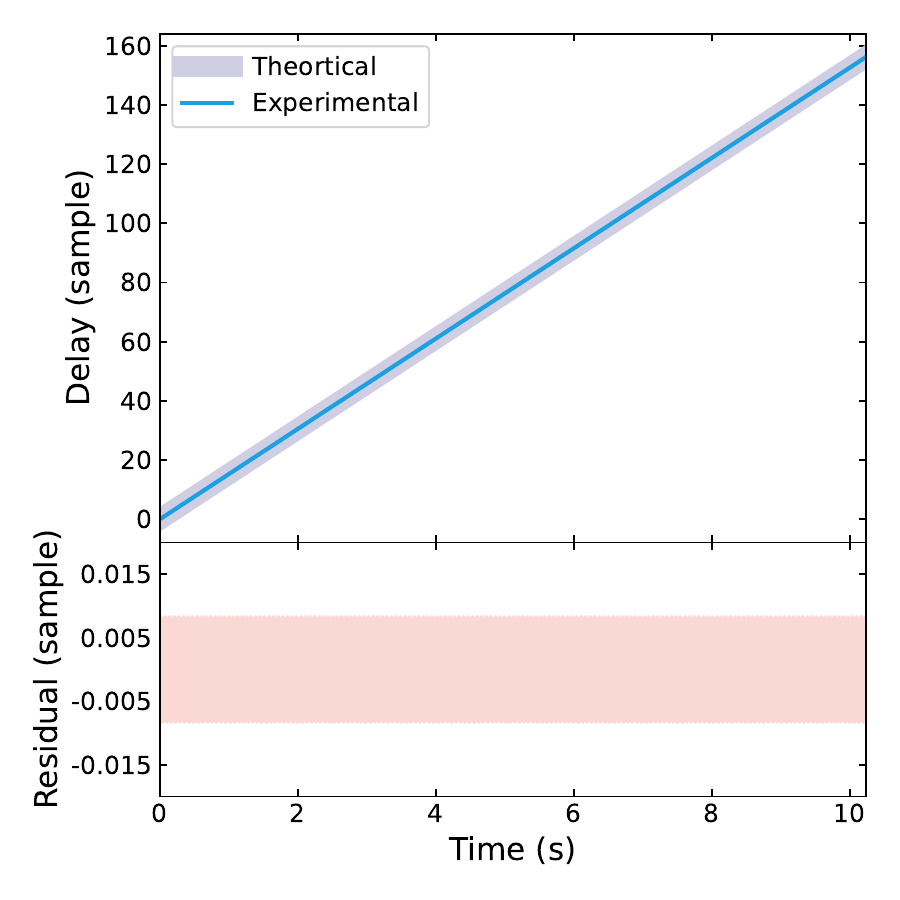}
    \caption{Measured delay based on the phase of the output data of the F-engine with a sine wave injection. The upper panel shows the measured delays (thin solid blue line) and injected delays (thick purple solid line). The bottom panel shows the delay residuals between measured delays and injected delays.}
    \label{fig:delay}
\end{figure}

To verify the delay correction, a synthesized 1 MHz  sine wave was injected into the data processing pipeline. A controlled delay increment was applied via the linear interpolation delay predictor to simulate a linearly increasing delay with time. The test setup involved receiving UDP packets containing the processed data over the network and storing them to the disk for offline analysis. This enabled direct observation of the F-engine output and evaluation of the delay correction performance. The primary diagnostic was the phase of the 1 MHz frequency channel output, i.e., the phase of the complex-valued FFT bin corresponding to that frequency. The observed phase shift was compared against the expectation of linearly changing phase with delay. The results, shown in \FIG{delay}, confirm that the system accurately compensates for the introduced delays: the measured phase shift closely tracks the theoretical expectation, corresponding to a delay of approximately 0.01 samples. This validates the correct functioning of the delay correction pipeline. A full-band complete delay test with noise will be performed using the correlator in the future. This requires multiple boards, since in our design each board handles one antenna, with the same geometric delay applied to all its inputs.

\subsection{Test observation}
\begin{figure}
    \centering
    \includegraphics[width=\linewidth]{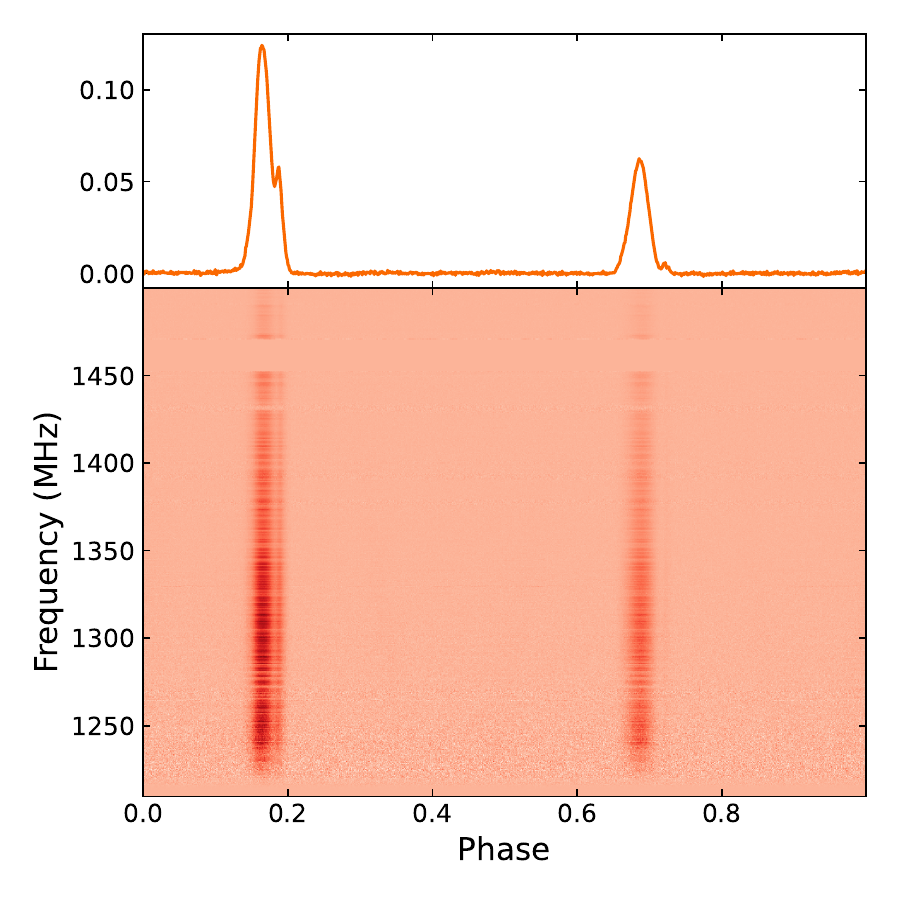}
    \caption{Dynamic spectrum of the millisecond pulsar PSR\,J1939+2154, obtained during the 2-hour test observation with the Effelsberg 100 m telescope.}
    \label{fig:dynamic_spectrum}
\end{figure}

\begin{figure}
    \centering
    \includegraphics[width=\linewidth]{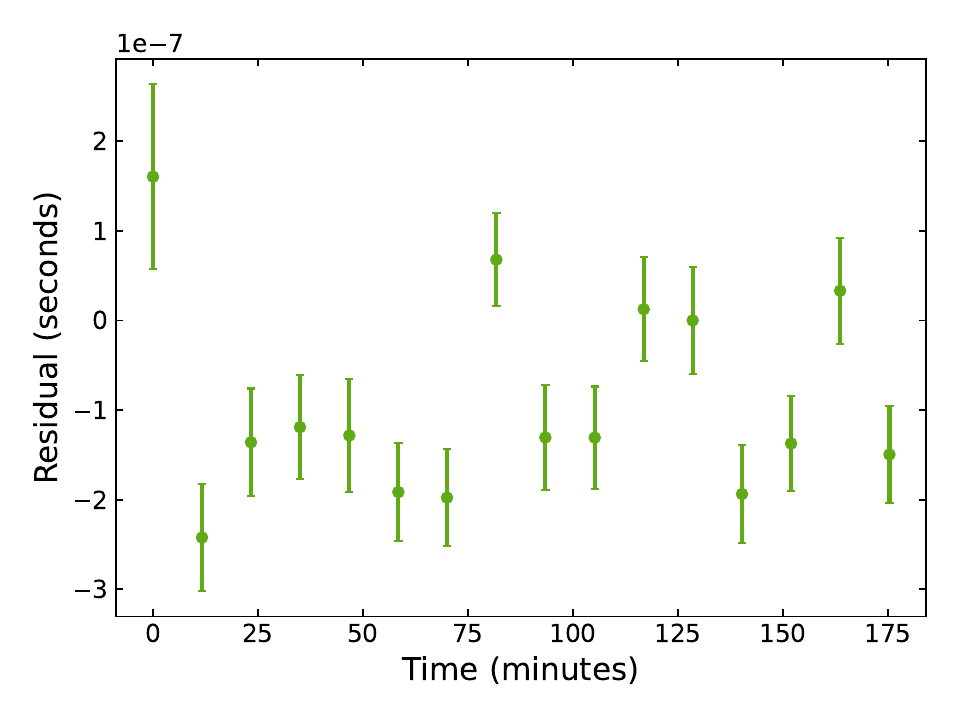}
    \caption{Timing residuals of the times  of arrival  (ToAs) from the millisecond pulsar PSR\,J1939+2154, obtained during the 2-hour test observation. Each ToA is formed in the session after rebooting the backend.}
    \label{fig:timing}
\end{figure}

To evaluate the practical performance of the F-engine, we conducted a test using the 100-meter Effelsberg radio telescope. We selected the millisecond pulsar PSR\,J1939+2154 as the test source due to its well-known microsecond-scale pulse structures and excellent short-term timing stability. A two-hour observation was carried out using the L-band receiver, covering a frequency range from 1200\,MHz to 1500\,MHz. To assess the timing robustness of the system, the F-engine was deliberately restarted multiple times during the observation. The output data were subsequently processed using coherent dedispersion and folding with {\sc DSPSR}.\footnote{https://dspsr.sourceforge.net/} The resulting dynamic spectrum of the full observation is shown in \FIG{dynamic_spectrum}, and the timing analysis is presented in \FIG{timing}. The fine temporal and spectral structures of PSR\,J1939+2154 are clearly visible in the dynamic spectrum, and the derived timing residuals exhibit a precision better than 1\,$\mu$s. These results validate the timing fidelity and overall practical performance of the F-engine in real observational conditions.

\section{Discussion}
\label{sec:Discussion}

\subsection{High-level synthesis}
In our F-engine design, the majority of modules were implemented using Xilinx Vitis HLS in C++, demonstrating the viability of high-level synthesis (HLS) for backend systems operating at high data rates in radio astronomy. HLS can significantly enhance FPGA development efficiency; however, its effective use requires a good understanding of FPGA architecture and the HLS toolchain. Without prior experience in hardware design, the learning curve can be steep, particularly since HLS inherits serial programming paradigms from conventional software development, which may not map efficiently to parallel hardware architectures.

Larger radio interferometers are becoming the norm in radio astronomy, offering wide fields of view, high sensitivity, and precise source localization. However, they generate massive data rates, making real-time processing essential due to the impracticality of storing raw data. Traditionally, FPGAs have been used for backend processing \citep{Men2019MNRAS}, while CPUs and GPUs are preferred for compute-intensive tasks such as pulsar and fast radio burst searches \citep{Ransom2002AJ, Men2024A&A}, owing to their ease of programming and high computational throughput. Nonetheless, FPGAs offer key advantages: deterministic low-latency processing, lower power consumption, and sustained high-throughput performance independent of PCIe bandwidth constraints. As HLS tools mature, the long-standing barrier of complex HDL-based development may be reduced, enabling the broader application of FPGAs in real-time astronomical data processing.

\subsection{Oversampled polyphase filter bank}
The critically sampled polyphase filterbank (PFB) is widely employed in radio telescope backends for channelization. However, the design of such PFBs requires a trade-off between the minimization of spectral leakage and the maximization of channel bandwidth \citep{Straten2011PASA}. Achieving wider channels while maintaining acceptable spectral leakage levels requires increasing the number of FIR filter taps, which in turn imposes additional resource demands on the FPGA. 

\begin{figure}
    \centering
    \includegraphics[width=\columnwidth]{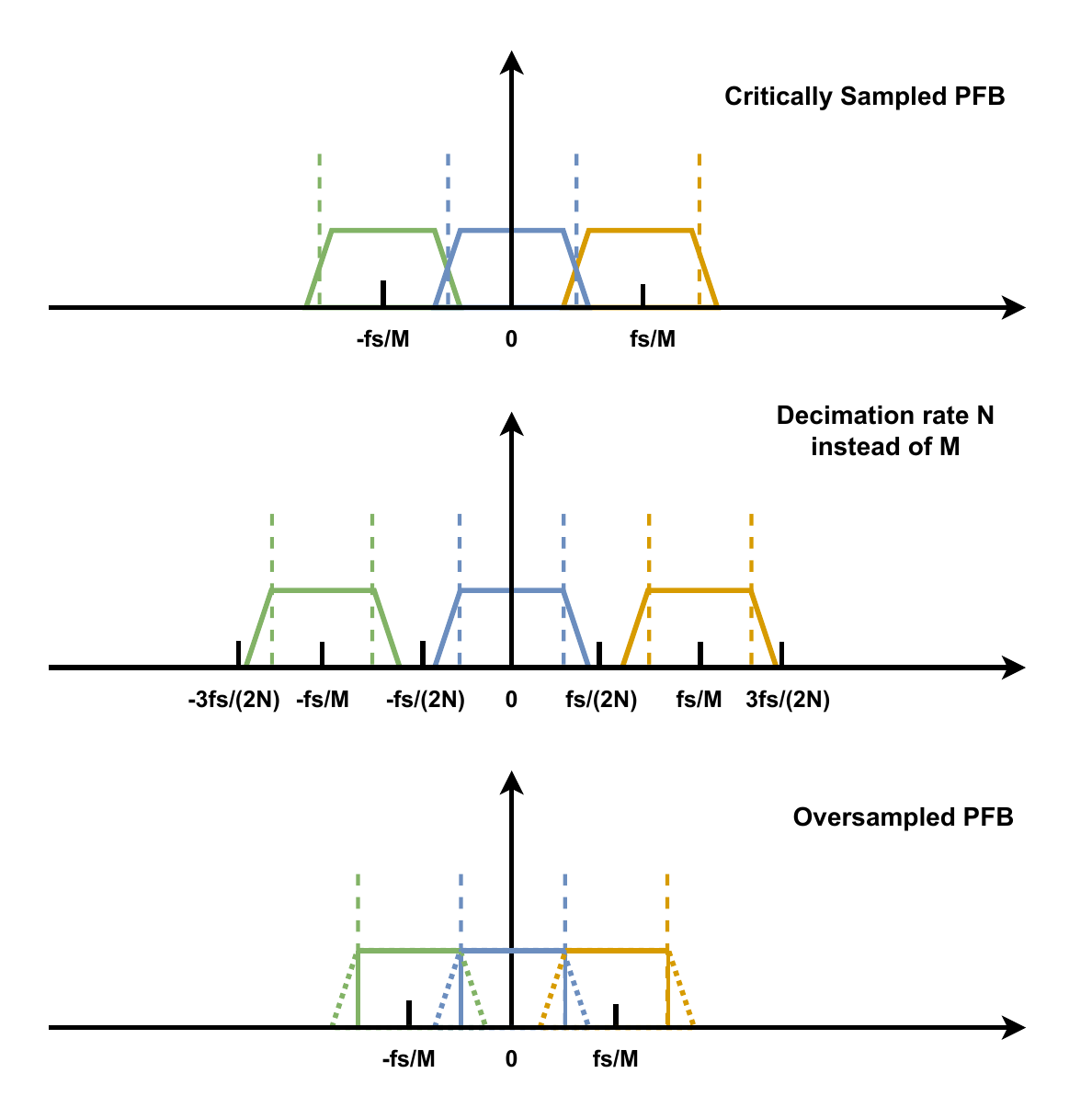}
    \caption{Upper panel: Channel response of a critically sampled PFB with spectral leakage. Middle panel: Channel response using a lower decimation rate of $N$ than the number of FFT points $M$, which can have asymmetric response with an offset between channel center and response center. Bottom panel: Channel response of an oversampled PFB after applying a second stage synthesis filter.}
    \label{fig:opfb}
\end{figure}

An alternative approach is the oversampled PFB technique \citep{HarrisIEEE2003}. This method improves spectral isolation by separating the filtering into two stages. First, a FIR filter followed by an FFT is applied as in the critically sampled case, but the output is oversampled, i.e., the decimation rate $N$ is less than the FFT length $M$, resulting in channel outputs with overlapping bandwidths. Second, a synthesized filter is then applied per channel to sharpen the effective cutoff response and further suppress spectral leakage. A comparison of the frequency responses between critically sampled and oversampled PFBs is shown in \FIG{opfb}.

However, when $N$ is not an integer divisor of $M$, the output channels are no longer centered at their nominal frequencies, introducing a frequency-dependent offset. This misalignment, also illustrated in \FIG{opfb}, can be corrected by applying a complex frequency shift (i.e., digital mixing) per channel, expressed as
\begin{align}
    y[k N, l] &= \sum^{M-1}_{m=0} \sum^{P-1}_{p=0} x[k N-(p M + m)] \cdot h[p M + m] \cdot \notag\\ &\exp \left(j 2\pi \frac{l}{M} (m - k N) \right)\,.
    \label{eq:opfb}
\end{align}
This additional phase term can be applied in the Fourier domain via complex multiplication, although at increased computational cost. Alternatively, it may be implemented in the time domain using a circular buffer approach, which can offer resource savings on FPGA platforms \citep{SmithIEEE2021}.

In our design, the HLS-based PFB implementation is parametrizable and supports reconfiguration to an oversampled PFB by modifying a small set of compile-time parameters. However, due to the increased data rate of the oversampled output, the design currently exceeds the available DDR4 bandwidth on the RFSoC4x2 board used for corner-turning, even though the network interface can accommodate the higher throughput. Future deployment on hardware with higher memory bandwidth would enable the straightforward adoption of the oversampled PFB without additional architectural changes.

\section{Conclusions}
\label{sec:Conclusions}
In this work we presented the development and implementation of the F-engine using the RFSoC4x2 board for the ARGOS project. The F-engine was designed to support critical functions such as channelization, delay correction, and complex gain calibration, which are essential for processing radio frequency signals in real-time astronomical observations. We utilized the Xilinx Vitis HLS C++ language to implement these modules, showcasing the effectiveness of HLS in developing high-performance backend systems for radio astronomy applications.

Through hardware verification, we validated the performance of the channel response and delay correction, demonstrating that the system achieved the expected theoretical performance. The ability of the F-engine to handle real-time processing tasks with high-precision highlights its potential for use in large-scale astronomical data processing. Additionally, we conducted pulsar observations using the F-engine at the Effelsberg 100-meter radio telescope, which served as a practical test and confirmed the functionality and performance of the system in an operational environment.

This work underscores the promising applications of HLS in high-performance computing, particularly in the context of astronomical data processing. Additionally, our work demonstrates an alternative to traditional CPU- and GPU-based approaches, offering the advantages of reduced power consumption, improved throughput, and the ability to scale for future data-intensive tasks. With the growing complexity of astronomical data and the increasing demand for real-time signal processing, FPGAs and HLS provide a compelling solution for the next generation of scientific computing.

\begin{acknowledgements}
This research was funded by Max Planck Society, the European Commission under the project ARGOS-CDS (Grant Agreement No. 101094354) and RADIOBLOCKS project (Grant Agreement No. 101093934). We are grateful to Stefan Heyminck for providing helpful suggestions.
\end{acknowledgements}

\bibliographystyle{aa}
\bibliography{ms}

\end{document}